\documentclass[12pt,osajnl,eqsecnum,preprint,showpacs]{revtex4}
\DeclareRobustCommand{\baselinestretch{2.2}}

\usepackage{amsfonts,amssymb,amsbsy,bm,subeqnarray,graphicx}

\begin{document}

\title{Characterizing the reflectance near
the Brewster angle: a Pad\'e-approximant approach}

\author{Teresa Yonte and
Luis L. S\'{a}nchez-Soto}

\affiliation{Departamento de \'Optica,
Facultad de F\'{\i}sica,
Universidad Complutense,
28040~Madrid, Spain}

\date{\today}

\begin{abstract}
We characterize the reflectance peak near the Brewster angle
for both an interface between two dielectric media and a
single slab. To approach this problem analytically, we
approximate the reflectance by a first-order diagonal Pad\'e.
In this way, we calculate the width and the skewness of the
peak and we show that, although they present a well-resolved
maximum, they are otherwise not so markedly dependent on the
refractive index.

\end{abstract}

\maketitle

\section{Introduction}

As it is well known, the behavior of the light reflected at the
boundary between two dielectrics drastically depends on whether
the associated electric field lies parallel ($p$ polarization)
or perpendicular ($s$ polarization) to the plane of incidence.
This is quantified by the Fresnel formulas~\cite{BornWolf1999},
which also confirm that at a certain angle  $\theta_B$ the
$p$-polarized field drops to zero. This angle is commonly referred
to as the Brewster angle and, in spite of its simplicity, it is
a crucial concept in the optics of reflection~\cite{Lekner1987}.

In fact, this notion is at the heart of a number of methods for
optical thin-film characterization. These include Brewster-angle
microscopy~\cite{Honig1991,Henon1991,Lheveder1998}, which
furnishes a direct visualization of the morphology of the film,
or the Abel\`es Brewster-angle method~\cite{Abeles1950,Traub1957,Hacskaylo1964,Heavens1991,Wu1993}
for determining the refractive index. They are quite popular
because are noncontact, easy to set up and simple to use. The
basic idea can be concisely stated as follows: coated and bare
regions of a substrate are illuminated with $p$-polarized light
and the angle of incidence is scanned for a reflectance-match
angle. Visual observation is often used because it reveals local
variations in film thickness and in homogeneity of refractive index.

These techniques rely on the dependence of reflectance near
the Brewster angle on the refractive indices involved. Of
course, one could rightly argue that all the relevant
information is contained in the Fresnel formulas. However,
given their nonlinear features, no physical insights on the
behavior around $\theta_B$ can be easily inferred, nor an
analysis of the error sources can be easily carried out, except
by numerical methods~\cite{Burns1974}. Some qualitative comments
can be found scattered in the literature: for example, it is
sometimes argued that the amplitude reflection coefficient for
$p$ polarization is approximately linear around $\theta_B$~\cite{Holmes1965}.
Nevertheless, we think that a comprehensive study of the reflectance
near the Brewster angle behavior is missing and it is precisely
the main goal of this paper.

To this end, we propose to look formally at the reflectance as a
probability distribution and focus on its central moments, which
leads us to introduce very natural measures of the width and the
skewness of that peak through the second and third moments~\cite{Evans2000}.
Since all these calculations must be performed only in a
neighborhood of $\theta_B$, we replace the exact reflectance
by a Pad\'e approximant~\cite{Baker1996}: apart from the elegance
of this approach, it is sometimes mysterious how well this can work.
In addition, we can then compute the relevant parameters in a closed
way and deduce their variation with the refractive indices. This
program is carried out in Section~\ref{inter} for a single interface
and in Section~\ref{layer} for a homogeneous slab. Finally, our
conclusions are summarized in Section~\ref{con}.

\section{Behavior of the Brewster angle at an interface}
\label{inter}

Let two homogeneous isotropic semi-infinite dielectric media,
described by real refractive indices $n_0$ and $n_1$, be
separated by a plane boundary. We assume an incident monochromatic,
linearly  polarized plane wave from medium 0, which makes
an angle $\theta_0$ with the normal to the interface and
has amplitude $E^{(i)}$. This wave splits into a reflected
wave $E^{(r)}$ in medium 0, and a transmitted wave $E^{(t)}$
in medium 1 that makes an angle $\theta_1$ with the normal.
The angles of incidence $\theta_0$ and refraction $\theta_1$
are related by Snell's law.

The wave vectors of all waves lie in the plane of incidence.
When the incident fields are $p$ or $s$ polarized, all plane
waves excited by the incident ones have the same polarization,
so both basic polarizations can be treated separately. By
demanding that the tangential components of $\mathbf{E}$
and $\mathbf{H}$ should be continuous across the boundary,
and assuming nonmagnetic media, the reflection and transmission
amplitudes are given by~\cite{BornWolf1999}
\begin{subeqnarray}
\label{rs}
r_s &  = &  \frac{E^{(r)}_s}
{E^{(i)}_s} =
\frac{n_0 \cos\theta_0 - n_1 \cos\theta_1}
{n_0 \cos\theta_0 + n_1 \cos \theta_1}, \\
& & \nonumber \\
\label{rp}
r_p & = &  \frac{E^{(r)}_p}
{E^{(i)}_p}   =
\frac{n_1 \cos\theta_0 - n_0 \cos\theta_1}
{n_1 \cos\theta_0 + n_0 \cos\theta_1} .
\end{subeqnarray}
These are the famous Fresnel formulas, represented in  Fig.~1
and whose physical content is discussed in any optics textbook.
The Brewster angle occurs when $r_p = 0$, which immediately
gives the condition
\begin{equation}
\theta_B =  \tan^{-1} n ,
\end{equation}
where $n = n_1 /n_0$ is the relative refractive index.
Without loss of generality, in the rest of this section
we assume $n_0 =1$ and then $n= n_1$. Also, we deal
exclusively  with $p$ polarization and drop the
corresponding subscript everywhere.

\begin{figure}
\centering
\resizebox{0.95\columnwidth}{!}{\includegraphics{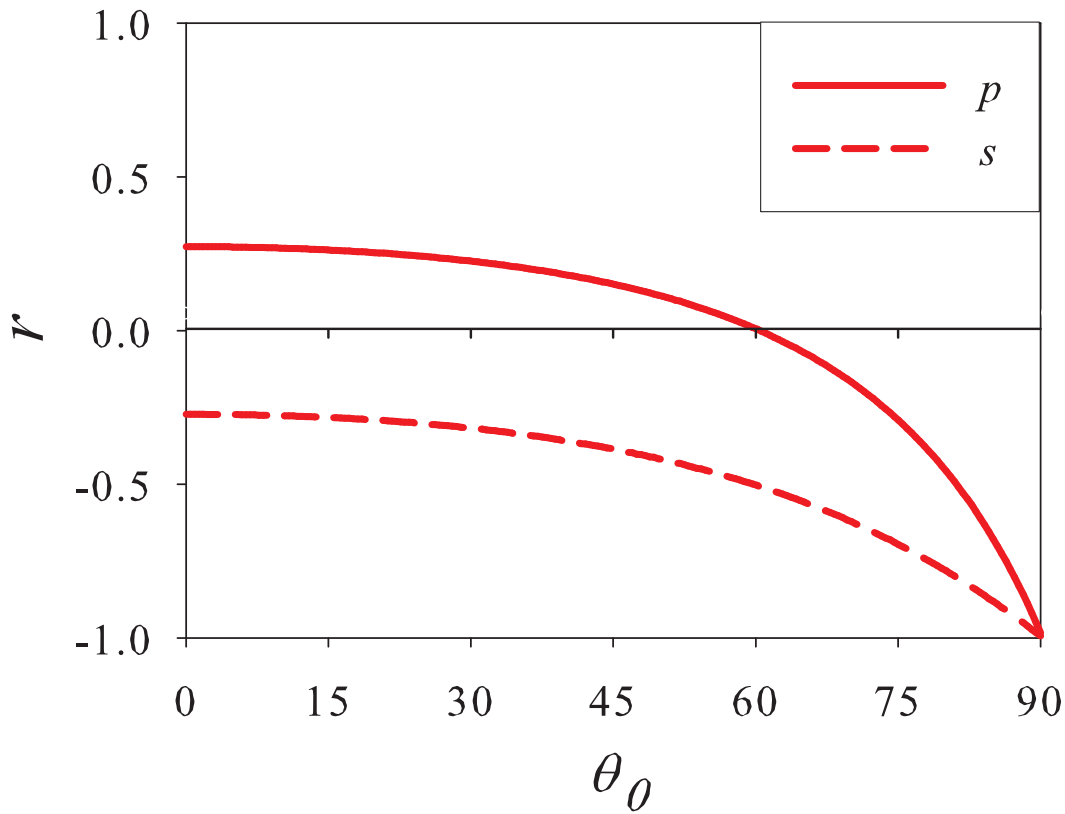}}
\caption{Plot of the amplitude reflection coefficients $r_p$
and $r_s$ as functions of the angle of incidence $\theta_0$
(in degrees) for an interface between air ($n_0 = 1$) and a
homogeneous, isotropic dielectric medium  ($n_1= 1.75$).}
\end{figure}

To treat the reflectance
\begin{equation}
\mathfrak{R}  = | r |^2
\end{equation}
near the Brewster angle it proves convenient to use
\begin{equation}
\label{locinc}
\vartheta = \theta_0 - \theta_B ,
\end{equation}
which is the angle of incidence centered at $\theta_B$. Since
we are interested only in a local study, we take into account
only a small interval around $\theta_B$; that is, angles
from $- \Delta \vartheta$ to $+ \Delta \vartheta$. The length
of this interval is largely an arbitrary matter: we shall
take henceforth $ \Delta \vartheta = 15^\circ$, although the
analysis is largely independent of this choice.

To quantify the peak near the Brewster angle, we treat the
reflectance as a probability distribution in the interval
$[- \Delta \vartheta, + \Delta \vartheta]$ and  borrow
some well-established concepts from statistics~\cite{Evans2000}.
In this manner we define
\begin{equation}
\bar{\vartheta} =
\frac{\int_{- \Delta \vartheta}^{+ \Delta \vartheta}
d\vartheta \, \vartheta \ \mathfrak{R} (\vartheta )}
{\int_{- \Delta \vartheta}^{+ \Delta \vartheta}
d\vartheta \,  \mathfrak{R} (\vartheta )} .
\end{equation}
For a symmetric peak $\bar{\vartheta} = 0$, while the fact
that $\bar{\vartheta} \neq 0$ reveals an intrinsic asymmetry.

The central moments are
\begin{equation}
\mu_k =
\frac{\int_{- \Delta \vartheta}^{+ \Delta \vartheta}
d\vartheta \, \left ( \vartheta - \bar{\vartheta} \right )^k \
\mathfrak{R} (\vartheta)}
{\int_{- \Delta \vartheta}^{+ \Delta \vartheta}
d\vartheta \,  \mathfrak{R} (\vartheta)} ,
\end{equation}
and, as it happens for $\bar{\vartheta}$, they are functions of
the refractive index $n$. The second moment is a measure of the
width of the distribution, while the third moment can be immediately
related with a lack of symmetry. More concretely, we take the
width and the skewness of the Brewster peak as
\begin{subeqnarray}
w & = & 2 \sqrt{\mu_2} , \\
& & \nonumber \\
\gamma & = &   \frac{\mu_3}{\sqrt{\mu_2^3}} .
\end{subeqnarray}

\begin{figure}
\centering
\resizebox{0.95\columnwidth}{!}{\includegraphics{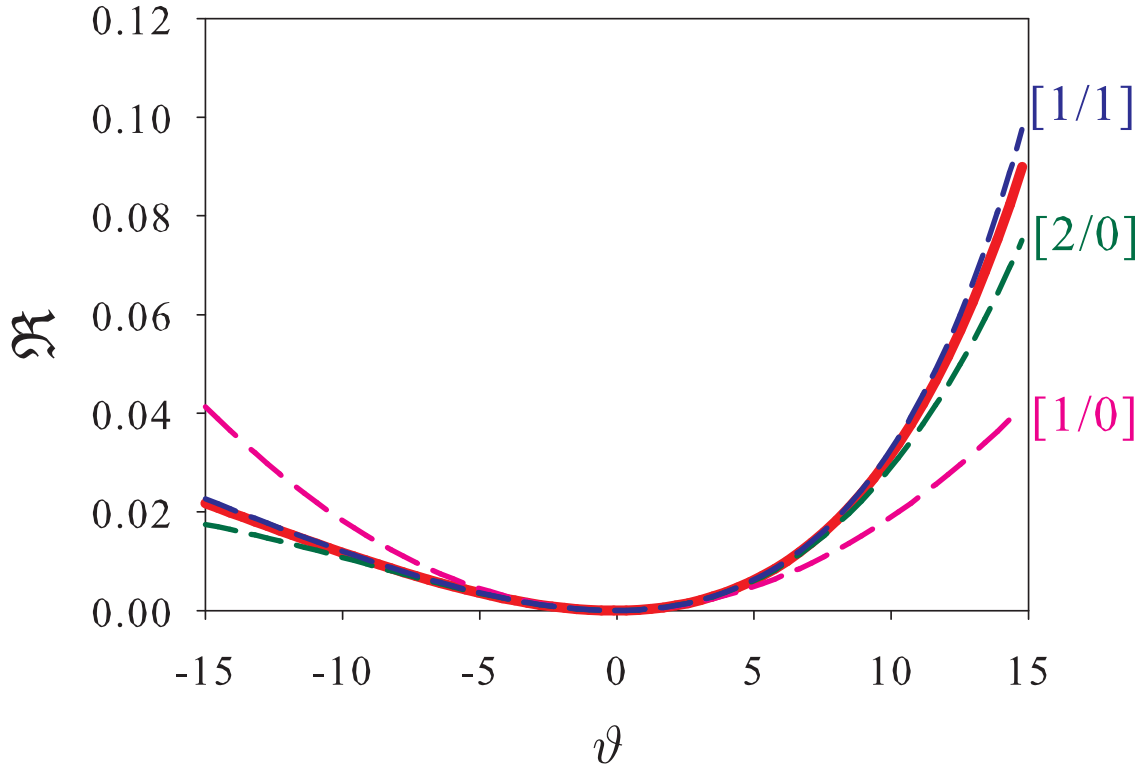}}
\caption{Reflectance (for $p$ polarization) of the same interface as
in Fig.~1, but calculated in terms of the local angle of incidence
$\vartheta$ (in degrees) for an interval $\pm \Delta \vartheta = 15^\circ$.
The exact reflectance is represented by the thick continuous line, while
the broken lines are the Pad\'e approximants in Eqs.~(\ref{PadeLMboun}).
The corresponding orders $[L/M]$ are labeled in the right axis.}
\end{figure}

Of course, these parameters can be computed numerically by using
the Fresnel formulas. However, since we are working in a small
neighborhood of $\theta_B$, it seems more convenient to work with
some local approximation to $r$ (and therefore to $\mathfrak{R}$).
Here, we resort to Pad\'e approximants (to maintain the paper as
self-contained as possible, in the Appendix we briefly recall
the essentials of such an approach). After some calculations,
the approximations to the reflection coefficient $r$ turn out to
be
\begin{subeqnarray}
\label{PadeLMboun}
r^{[1/0]} &  = &  \frac { (1 - n^4) \vartheta}{2 n^3} , \\
r^{[2/0]} & = & \frac{(1 - n^4) \vartheta}{2 n^3} \nonumber \\
& - & \frac{( n^8 + n^6 + n^4 - n^2 - 2 ) \vartheta^2}
{4n^6} , \\
r^{[1/1]} & = & \frac{(1- n^4) \vartheta }
{2 n^3 - (n^4 + n^2 + 2) \vartheta} ,
\end{subeqnarray}
whence one immediately gets the corresponding reflectances
as $\mathfrak{R}^{[L/M]} = |r^{[L/M]}|^2$, which have been
plotted in Fig.~2. As we can see, the term $\mathfrak{R}^{[1/0]}$
reproduces the exact reflectance only in a very small interval.
The next polynomial approximation $\mathfrak{R}^{[2/0]}$ improves
a little bit the situation, but fails again as soon we are,
say $\pm 10^\circ$ away from $\theta_B$. On the contrary,
the diagonal Pad\'e $\mathfrak{R}^{[1/1]}$ fits remarkably
well with the exact behavior. We thus conclude that
$\mathfrak{R}^{[1/1]}$ provides an excellent approximation
and we use it in subsequent calculations.

\begin{figure}
\centering
\resizebox{0.95\columnwidth}{!}{\includegraphics{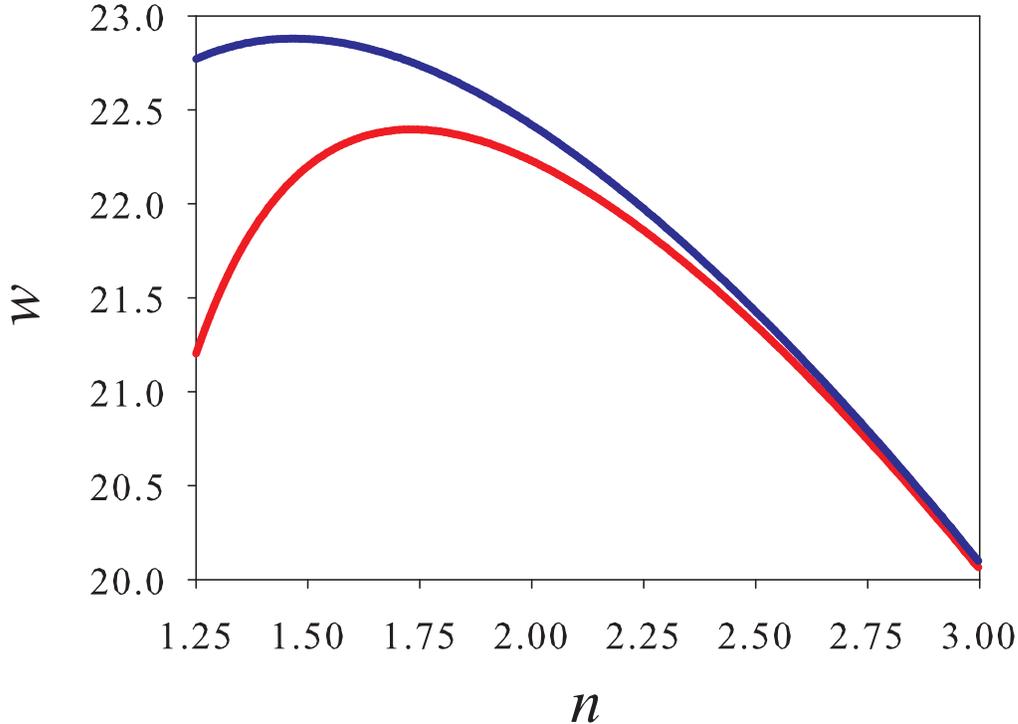}}
\caption{Width (in degrees) of the reflectance peak near the
Brewster angle as a function of the refractive index $n$ for
the same interface as in Fig.~1 (lower curve) and for a slab
of index $n$ and thickness $d = \lambda/4$ (upper curve).}
\end{figure}

An additional and remarkable advantage of $\mathfrak{R}^{[1/1]}$
is that the central moments $\mu_k$ can be expressed in a closed
analytical form.  In Fig.~3 we have plotted the width $w$ as a
function of the refractive index $n$ (lower curve). We have also
computed $w$ by numerically integrating the Fresnel formulas:
no appreciable differences can be noticed. The width has a maximum
that can be calculated by imposing $dw/dn = 0$, which immediately
gives $n = \sqrt{3}$. Beyond this value, $w$ decreases almost
linearly with $n$. However, for the range of indices plotted in
the figure, this variation is smooth, with a total change in $w$
of around $3^\circ$.

\begin{figure}
\centering
\resizebox{0.95\columnwidth}{!}{\includegraphics{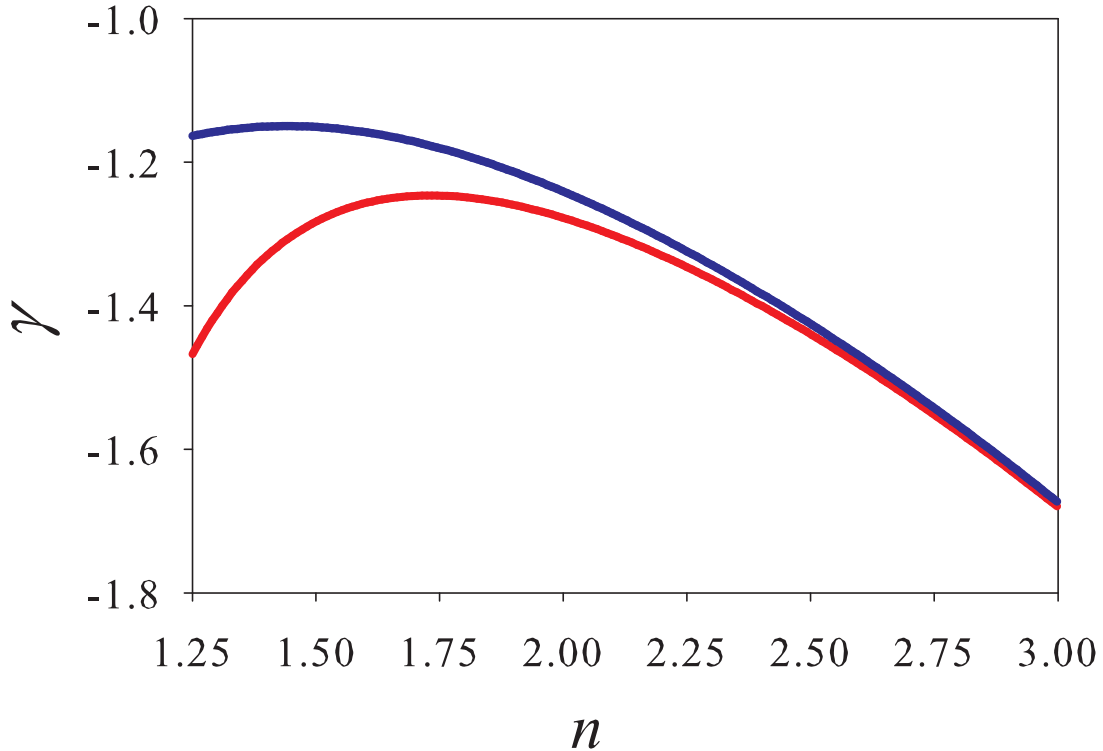}}
\caption{Skewness $\gamma$ of the reflectance peak near the
Brewster angle as a function of the refractive index, for the
same cases as in Fig.~3. The lower curve also corresponds to the
interface, while the upper one if for the slab.}
\end{figure}

In Fig.~4 we have plotted the skewness $\gamma$ in terms of $n$.
This parameter is always negative what, for a peak, means that
the left tail is less pronounced than the right tail. There is
again a maximum in the skewness that can be calculated by imposing
$d\gamma / dn= 0$. However, one can check that $\mu_3 \simeq 1$ for
all the values of $n$, so that $\gamma \simeq - 8/w^3$: this
maximum coincides then with that of $w$. Apart from a scale
factor, $\gamma$ shows an aspect quite similar to that of  $w$.

\section{Behavior of the Brewster angle at a single slab}
\label{layer}

We focus our attention now on a homogeneous, isotropic dielectric
slab of refractive index $n$ and thickness $d$ imbedded in air
and illuminated with  monochromatic light  (the case when the
slab lies on a substrate can be dealt with much in the same way).
A standard calculation gives the following amplitude reflection
coefficient for the slab~\cite{Azzam1987}
\begin{equation}
\label{Rlam}
R = \frac{r [1 - \exp(- i 2 \beta ) ]}
{1- r^2  \exp(- i 2 \beta ) } ,
\nonumber
\end{equation}
where $r$ is the Fresnel reflection coefficient at the interface
air-medium (which again will be considered for $p$ polarization only)
and $\beta$ is the slab phase thickness
\begin{equation}
\beta = \frac{2 \pi}{\lambda} d \sqrt{n^2 - \sin^2 \theta_0} .
\end{equation}
Here $d$ is the thickness, $\lambda$ the wavelength in vacuo
of the incident radiation and $\theta_0$ the angle of incidence.
This coefficient $R$ presents a typical periodic variation
with the slab phase thickness $\beta$. Apart from that dependence,
it is obvious that $R=0$ when $r=0$; i. e., precisely at the
Brewster angle for the interface air-medium. However, the form
of this reflectance peak, calculated now as
\begin{equation}
\mathfrak{R} = | R |^2 ,
\end{equation}
differs from the case of a single interface, since (\ref{Rlam})
is more involved than (\ref{rp}). Although the analysis can
be carried out for any value of $d$ such that $R \neq 0$, for
definiteness and to avoid as much as possible any spurious
detail, we take $d=\lambda /4$. In Fig.~5 we have plotted
the exact form of the reflectance for this thickness, as a
function of the local angle of incidence, defined as in
Eq.~(\ref{locinc}), for several values of $n$

\begin{figure}
\centering
\resizebox{0.95\columnwidth}{!}{\includegraphics{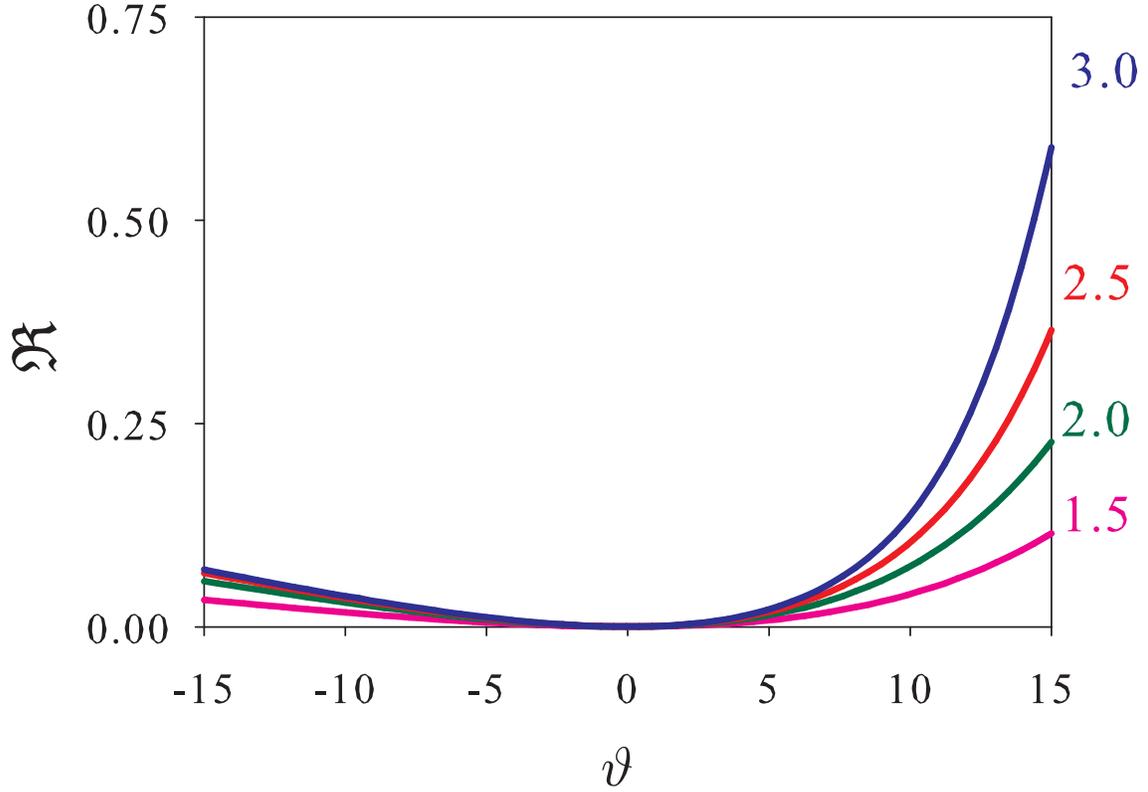}}
\caption{Reflectance (for $p$ polarization) of a slab of thickness
$d = \lambda/4$ in terms of the local angle of incidence $\vartheta$,
for various values of the refractive index (labeled in the right axis).}
\end{figure}

Instead of Eq.~(\ref{Rlam}), we use again a Pad\'e approximant
of order $[1/1]$, which can be written as
\begin{equation}
R^{[1/1]} = \frac{p_1 \vartheta}{1 + q_1 \vartheta}
\end{equation}
where
\begin{subeqnarray}
p_1 & = &  i \frac{1- n^4}{n^3} e^{-i \nu} \sin \nu  ,   \\
& & \nonumber \\
q_1 & = & \frac{\nu}{n^3 e^{i \nu} \sin \nu} -
\frac{n^4 + n^2 +2}{2 n^3} ,
\end{subeqnarray}
and
\begin{equation}
\nu = \frac{2\pi n^2 d}{\sqrt{n^2+1}} .
\end{equation}
With this approximation, that still reproduces pretty well the
exact reflectance, we have analytically obtained  $w$ and
$\gamma$ for the slab. The results are plotted in the upper
curves of Figs.~3 and 4, respectively. For high values of $n$,
the reflectance peak for the slab essentially coincides
with that for the interface. Differences are only noticeable
for small values of $n$. In fact, the maximum of $w$ can be
evaluated by imposing $dw/dn = 0$, whose numerical solution
gives $n \simeq 1.45$. The behavior of $\gamma$ is quite similar,
because here we have also that $\mu_3 \simeq -1$ for all the values
of $n$.

\section{Conclusions}
\label{con}

In summary, we have presented a simple and comprehensive treatment
of the reflectance near the Brewster angle. By combining the
notions of width and skewness with Pad\'e approximants, we
have fully characterized this reflectance peak. We hope that
these results fill a gap for a full understanding of the Brewster angle.

\section*{Acknowledgments}
We thank Juan J. Monz\'on for many inspiring discussions and for
a careful reading of the manuscript.

\appendix

\section*{Appendix: Pad\'e approximants}

\setcounter{equation}{0}
\renewcommand{\theequation}{A{\arabic{equation}}}

A Pad\'e approximant is a rational function (of a specified order)
such that its power series expansion agrees with a given power
series to the highest possible order. In other words, let us define
\begin{equation}
R^{[L/M]} (x) = \frac{P_L (x)}{Q_M (x)} ,
\end{equation}
where $P_L (x)$ and $Q_M (x)$ are polynomials of degrees $L$ and $M$,
respectively:
\begin{subeqnarray}
P_L (x) & = & p_0 + p_1 x + \ldots + p_L x^L , \\
& & \nonumber  \\
Q_M (x) & = & 1 + q_1 x + \ldots + q_M x^M .
\end{subeqnarray}
The rational function $R^{[L/M]} (x)$ is said to be a Pad\'e
approximant to the function $f(x)$, which has a Taylor expansion
at $x=0$
\begin{equation}
f(x) = \sum_{k=0}^\infty f_k \ x^k ,
\end{equation}
if
\begin{subeqnarray}
\label{propPad}
& f(0) = R^{[L/M]}(0) , &   \\
& & \nonumber \\
& \displaystyle
\left .
\frac{d^k}{dx^k} f(x) \right |_{x=0} =
\left .
\frac{d^k}{dx^k} R^{[L/M]} (x) \right |_{x=0} ,
\end{subeqnarray}
for $k= 1, 2, \ldots, L + M $. These conditions furnish $L + M + 1$
equations for $p_0, \ldots, p_L$ and $q_1, \ldots, q_M$. The
coefficients can be found by noticing that Eqs.~(\ref{propPad})
are equivalent to
\begin{equation}
f(x)-\frac{P_L (x)}{Q_M (x)} = 0 ,
\end{equation}
up to terms of order $L + M +1$. This gives directly the following
set of equations:
\begin{eqnarray}
\label{defeqPad}
& & f_0  = p_0 , \nonumber \\
& & f_1 + f_0 q_1 = p_1 , \nonumber \\
& & f_2 + f_1 q_1 + f_0 q_2 = p_2 , \nonumber \\
& &
\; \vdots \nonumber \\
& & f_L + f_{L-1} q_1 + \ldots + f_{L-M+1} q_M = 0 , \nonumber \\
& &
\; \vdots \nonumber \\
& & f_L + f_{L-1} q_1 + \ldots + f_{L-M+1} q_M = 0 , \nonumber \\
& & q_{L+M}  + f_{L + M -1} q_1 + \ldots + f_L q_M = 0 .
\end{eqnarray}
Solving this directly gives
\begin{widetext}
\begin{equation}
R^{[L/M]} (x) =
\frac{
\left |
\begin{array}{cccc}
f_{L-M+1} & f_{L-M+2} & \ldots & f_{L+1} \\
\vdots &  \vdots & \ddots & \vdots \\
f_{L} & f_{L + 1} & \ldots & f_{L+M} \\
\displaystyle
\sum_{J=M}^L f_{J-M} \ x^J &
\displaystyle
\sum_{J=M-1}^L f_{J-M+1}\ x^J &
\ldots &
\displaystyle
\sum_{J=0}^L f_J \ x^J
\end{array}
\right |}
{\left |
\begin{array}{cccc}
f_{L-M+1} & f_{L-M+2} & \ldots & f_{L+1} \\
\vdots &  \vdots & \ddots & \vdots \\
f_{L} & f_{L+1} & \ldots & f_{L+M} \\
x^M & x^{M-1} & \ldots & 1
\end{array}
\right |} .
\end{equation}
\end{widetext}
Nevertheless, experience shows that Eqs.~(\ref{defeqPad}) are
frequently close to singular, so that one should solve them
by e. g. a full pivotal lower-upper triangular (LU)
decomposition~\cite{Numericalrecipes}.

By contrast with techniques like Chebyshev approximation or
economization of power series, that only condense the information
that you already know about a function, Pad\'e approximants can
give you genuinely new information about your function
values~\cite{Numericalrecipes}. We conclude by noting that, for
a fixed value of $L + M$, the error is usually smallest when
$L = M$ or when $L=M+1$.

\end{document}